\documentclass{article}

\usepackage{arxiv}

\usepackage[utf8]{inputenc} % allow utf-8 input
\usepackage[T1]{fontenc}    % use 8-bit T1 fonts
\usepackage{hyperref}       % hyperlinks
\usepackage{url}            % simple URL typesetting
\usepackage{booktabs}       % professional-quality tables
\usepackage{amsfonts}       % blackboard math symbols
\usepackage{nicefrac}       % compact symbols for 1/2, etc.
\usepackage{microtype}      % microtypography
\usepackage{lipsum}
\usepackage{graphicx}
\usepackage{pifont}% http://ctan.org/pkg/pifont
\usepackage{tabularx}
\newcolumntype{L}[1]{>{\raggedright\let\newline\\\arraybackslash\hspace{0pt}}m{#1}}
\newcolumntype{C}[1]{>{\centering\let\newline\\\arraybackslash\hspace{0pt}}m{#1}}
\newcolumntype{R}[1]{>{\raggedleft\let\newline\\\arraybackslash\hspace{0pt}}m{#1}}

\title{RRD-Bio: Building An Integrated Research Resource Database for Biomedicine}

\author{
 \href{https://orcid.org/0000-0003-2104-0194}{Li Zhang \hspace{1mm} \includegraphics[scale=0.08]{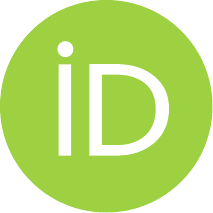}} \\
  School of Information Management\\
  Nanjing University\\
163 Xianlin Avenue, Qixia District, Nanjing, Jiangsu Province, China \\
  \texttt{zlahu@foxmail.com} \\
     \And
 Mengting Sun \\
  School of Information Management\\
  Wuhan University\\
  No. 299 Bayi Road, Wuchang District, Wuhan, Hubei Province, China \\
  \texttt{ida\_sun@foxmail.com} \\
   \And
 Chong Jiang \\
  School of Information Management\\
  Wuhan University\\
  No. 299 Bayi Road, Wuchang District, Wuhan, Hubei Province, China \\
  \texttt{jiangc\_prayer@whu.edu.cn} \\
  \And
 Haihua Chen \\
  Department of Information Science\\
  University of North Texas\\
  Discovery Park, Denton, TX, 76201, US \\
  \texttt{Haihua.Chen@unt.edu} \\
}
 
\begin{document}
\maketitle
\begin{abstract}
Research resources (RRs) such as data, software, and tools are essential pillars of scientific research. The field of biomedicine, a critical scientific discipline, is witnessing a surge in research publications resulting in the accumulation of a substantial number of RRs. However, these resources are dispersed among various biomedical articles and can be challenging to locate and reuse due to their transient nature.
In this paper, we report our recent progress of biomedical data curation - building a large research resource database for biomedicine (RRD-Bio), based on a collection of 40 million papers from two large biomedical literature databases, PubMed and PubMed Central. The database contains 2,555,116 RRs, each identified by a location on the Internet (URL) and descriptive information (Context). We made the RRD-Bio database publicly available (\url{https://zenodo.org/records/10526493}) to enhance the visibility of biomedical research resources, the ability to preserve important resources and the reproducibility of biomedical research.
\end{abstract}

% keywords can be removed
\keywords{Research Resource \and URL \and Biomedical article}

\section{Introduction}

Scientific breakthroughs depend on the exploration of research questions. This process relies heavily on using research resources (RRs), such as tools and data, to analyze research material in depth and gain scientific discoveries.
in recent decades, benefiting from the rapid development of information technology and the Internet, a large number of research resources, which could only be reached through physical media, have gradually been digitized and appeared on the Internet, where they can be accessed and cited through a unified mechanism - Unified Resource Locators (URLs). The significant advancement greatly facilitates the dissemination of research resources and promotes the use of RR in research.

Due to the easy accessibility, using electronic RRs has become increasingly popular in many research fields and gradually become a general research paradigm. Biology is such a field that heavily uses electronic RRs. Researchers in this field focus on health and disease treatment issues, use resources like antibodies, model organisms, and software projects to explore research questions and cite the resources in the literature to report scientific innovations.

With the proliferation of biomedical research and many papers published, the field has amassed a large number of research resources. These RRs undoubtedly drive the advancement of biomedical research. However, such a large number of RRs also poses in-trivial challenges to researchers. Specifically, (1) These resources are scattered throughout numerous research articles, making it difficult for researchers to know what RRs are available. (2) The large number of research resources has left researchers in an information overload situation, making it difficult to accurately access resources that are of interest to researchers, especially for beginners. (3) If these problems are not addressed properly, they may exacerbate the reproducibility crisis and further hinder the dissemination of valuable research discoveries. A prior study pointed out that the identification of research resources is a serious problem for reproducibility because, in many studies, the organism strains used in the experiments or the use of particular antibodies could not be identified\cite{vasilevsky2013reproducibility}.

Similar to other types of electronic resources, biomedical research resources may also undergo construction, evolution, and extinction processes, these features of RR increase the complexity of the above issues. Moreover, with more and more biomedical research being published, the problems will be further strengthened. Therefore, there is a need to identify and manage biomedical research resources effectively.

To address this issue, our study establishes a large Research Resource Database for biomedicine (RRD-Bio). The database is created based on an important observation that most research resources are explicitly and properly cited in the literature, which is also a formalized scientific writing practice. This observation demonstrates that building RRD-Bio from published biomedical articles is feasible. Here we built our RRD-Bio database based on two large open-access biomedical databases, PubMed and PubMed Central (PMC), which contain over 30 million abstracts and 4 million full-text articles. Since these two databases cover almost all global biomedical articles, the resource database constructed here will capture nearly all research resources in this field.

%cover all important research resources in the field.
A large and comprehensive research resource database can help researchers quickly find the target RRs without having to search and read full-text articles, alleviate their difficulties in accessing scholarly resources among a vast array of biomedical articles, improve the efficiency of scientific research, and benefit the development of biomedical innovations.
In this article, we report our progress in constructing the RRD-Bio database, including the background, the building method, and some statistical analyses of the database. Additionally, we show a serving framework of RRD-Bio, which aims to maximize the value of the database by enhancing the processing of raw data and establishing an essential service system.

\section{Related works}
Research resources are the foundation of research and have received extensive attention from various research fields. In 2013, a neuroscience conference paid special attention to this issue, an important discussed topic was how to identify research materials such as antibodies used in neuroscience papers with machine learning to enhance the reproducibility and traceability of resource utilization. Subsequently, the identification of RR has drawn increasing attention. A landmark event is the creation of Research Resource Identifiers (RRIDs) in 2016, which advocate for the identification, discovery, and reuse of research resources with a unique ID to assist researchers in citing essential resources.

Advocated by the initiative, the Resource Identification Portal\footnote{https://www.rrids.org/} was established. The portal serves as a centralized platform for accessing and exploring Research Resource Identifiers. Afterward, a similar research resource portal SciCrunch\footnote{https://scicrunch.org} that supports the RRID proposal was established. In addition to these RRID-compatible platforms, the research community also established many research data repositories, such as DataCite\footnote{https://support.datacite.org/docs/getting-started}, and figshare\footnote{https://figshare.com/}, which different from the above portals, focus mainly on research data sharing and archiving.

In contrast to these endeavors, there are also many studies focusing on extracting research resources from papers, which is similar to the goal of this paper. However, the content extracted from these studies is less relevant to the resource. For example, many studies focused on extracting scientific entities from articles, such as diseases, proteins, and genes\cite{fei2021enriching}, and some studies focused on extracting citations and citation contexts\cite{peroni2020opencitations}. Compared to the above studies, the one that is closer to ours is \cite{wang2020microsoft}, this study offers extracted research resources on a well-known scholarly database, Microsoft Academic Graph (MAG). However, their resources focus on only research datasets and code, and the extracted resources are limited in size, containing only 65797 resources out of over 200 million papers.

\section{Method}

\begin{figure}
\centering
\includegraphics[width=0.7\linewidth]{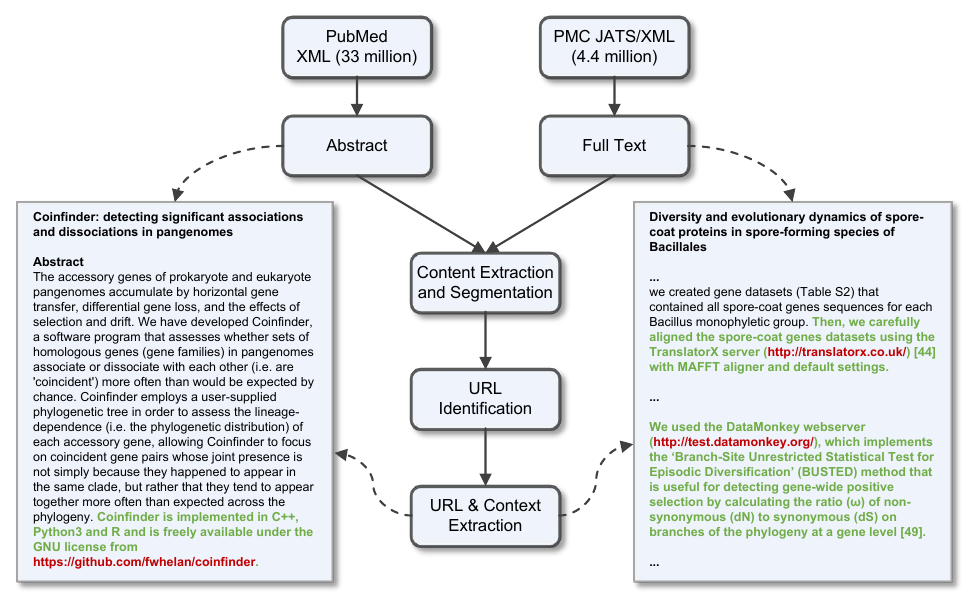}
\caption{The method of building RRD-Bio}
\label{fig:method}
\end{figure}

We followed the method illustrated in Fig.~\ref{fig:method} to build the RRD-Bio database. The research resource we extracted consists of two fields: resource link (URL) and resource context. The context field represents the contextual information surrounding the URLs in the paper. The reason that we extract this field is the resource link only conveys limited information about the resource, whereas the context field contains richer information about the resource such as resource name, creator, type, features, functionalities, frequent research questions associated with this resource, and instructions of use. By integrating the two fields, the extracted RR helps to enhance researchers' understanding of the resources and, therefore will contribute to the development of effective resource retrieval and recommendation systems\cite{patra2020content}.

% Note the method of building the database

To build RRD-Bio, we obtained the PubMed and PMC databases from NCBI's official site\footnote{https://ftp.ncbi.nlm.nih.gov/}. The baseline versions of the two databases are June 2022 for PMC and January 2022 for PubMed, containing 33 million abstracts and 4 million full-text articles, respectively.
Since these two databases are encoded in different XML formats, we developed two sets of codes to extract all textual content of research articles. Note that, the PMC articles are encoded loosely, which results in the URL links being encoded with many patterns\footnote{PMC article data are submitted by various articles providers (e.g., publishers), the same article metadata may be encapsulated by many XML tags, thus the XML format of PMC articles show a significant diverse, though they are all compliant with the JATS standards}. For example, URL links can be as a part of sentences, or as a footnote referring to a sentence. To handle this, We carefully examined all possible situations and made our extraction code accordingly.

% Note sentence tokenization
Then we tokenize the extracted content to sentences using a popular NLP toolkit NLTK. This step aims to split the whole texture context into many sentence pieces so that the research resource can be easily found in the pieces. After applying this tool to all PubMed and PMC papers, we obtained more than 600 million sentences.

% Note extraction strategy: regular expression
% note resource context
%
Based on the tokenized sentences, we used the following regex expression \verb|[a-zA-z]+://[\^\\s]*| to detect and extract research resources, which can handle various styles of URL links, such as links starting with ``http://'', or ``ftp://''. The rule for research resource judgment is that a sentence is considered as a research resource if it contains a valid URL. If contained, the words surrounding the URL are regarded as resource context.

\section{Results}

\subsection{Basic statistics of RRD-Bio}
% Note database size, PubMed/PMC, sentence length, number of citations, domains, over times.
% Note the number of (uniq) URLS each paper.
The RRD-Bio database contains 2555116 biomedical resource mentions, of which 95.8\% were collected from PMC. Since RRD-Bio was built on top of two scholarly databases, we show several key statistics of RRD-Bio resource instances according to the originated database. The results are shown in Table \ref{tab:basic-statistics}.

\begin{table*}
\centering
\caption{Basic statistics of RRD-Bio}
\label{tab:basic-statistics}
\begin{tabular}{lrr}
\toprule
\textbf{}                                       & \textbf{PubMed}          & \textbf{PMC}         \\
\midrule
Number of Papers                                         & 22,554,545             & 4,111,770              \\
Number and percentage of papers mentioning resource URLs & 94,352 (0.42\%)       & 766,945 (18.65\%)     \\
Number of Resources                                      & 107,544               & 2,447,572              \\
Averaged number of resource URLs per paper               & 1.14          & 3.19         \\
Averaged number of resource URLs per token & $2.44\times 10^{-5}$ & $1.70\times 10^{-4}$ \\
Number and Percentage of unique resource URLs            & 72,243 (61.17\%)      & 996,985 (40.73\%)     \\
%        Number of unique resource URLs1                          & 72,243                 & 785,981               \\
%        Number of unique resource URLs domains                & 24,198                 & 137,294               \\
%        avg num terms in url context      & 287.7005412           & 23.84637755          \\
\bottomrule
\end{tabular}
\end{table*}

We find that, in the PubMed and PMC databases, there are 33,434,749 and 4,441,815 articles, among which, 22,554,545 and 4,111,770 articles have non-empty content. In these non-empty articles, 0.42\% of PubMed articles and 18.65\% of PMC articles contain URL links. The two scholarly databases collectively contribute to 2,555,116 resource mentions, with 95.8\% contributed by PMC (2,447,572 vs. 107,544), indicating that the PMC database is the main RR contributor. This is not surprising as PMC is a full-text database, providing more complete
scientific articles than PubMed, despite PubMed having a higher quantity of papers.
The second finding is that the average number of resource mentions in PMC articles is 3.19, 2.8 times higher than in PubMed articles. In fact, in addition to the longer content of articles, PMC tends to contribute more resources at the token level. As indicated by the ``Averaged number of resource URLs per token'' in Table \ref{tab:basic-statistics}, PMC articles have a higher frequency or density of resource mentions. This disparity can be attributed to the fact that resource links in PubMed papers often refer to the resources created by researchers, such as the GitHub code repository, while in PMC, the resources encompass not only the researcher-created resources but also a substantial percentage of resources created by others. The ``Number and Percentage of unique resource URLs'' in Table \ref{tab:basic-statistics} justifies this inference, it shows that 40.73\% of resources in PMC are unique, whereas in PubMed, unique resources account for 61.17\%. Third, we observed that only 42\% resource links in PubMed can be found in PMC, highlighting the differences in resource mention patterns between the two databases.

\subsection{Resource popularity Analysis}

We analyze the popularity of research resources from two dimensions: URL-Freq and Domain-Freq, representing the frequency of URLs and the frequency of URL domains in our RRD-Bio database, respectively. The URL domain reflects the providers of resources. For example, the domain of the URL \url{http://trendscenter.org/software/gift/} is \url{http://trendscenter.org}. The domain analysis groups resources of the same providers into categories, it helps understand the resource providers in biomedicine and depicts the distribution of biomedical resources on the web.

\begin{figure}
\centering
\includegraphics[width=0.7\linewidth]{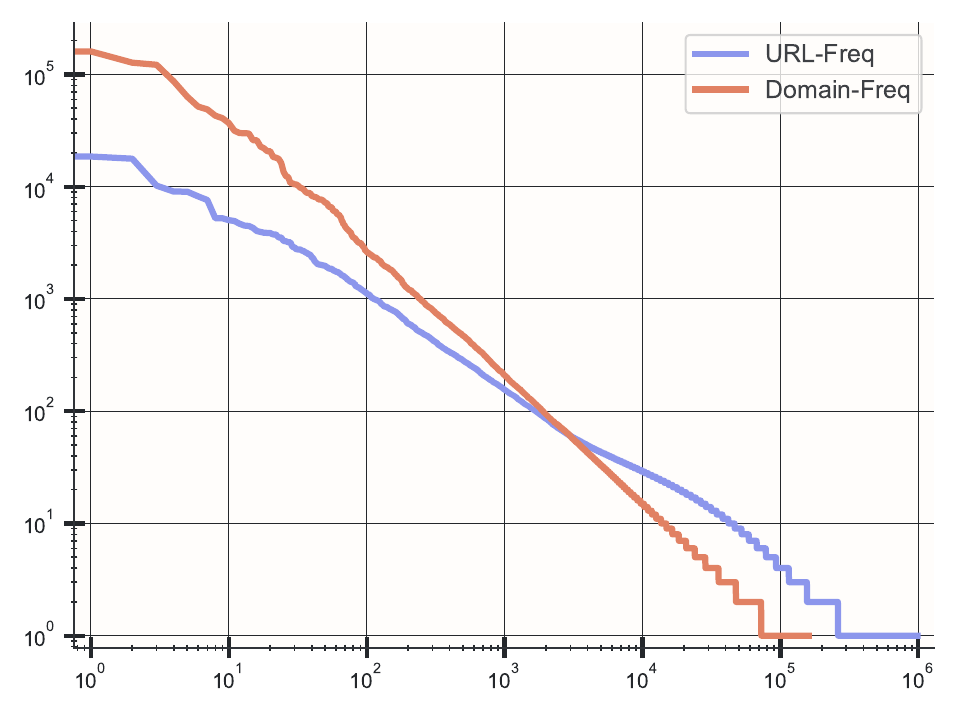}
\caption{Resource hotness}
\label{fig:resource-hotness}
\end{figure}

Figure \ref{fig:resource-hotness} illustrates the distribution of URL-Freq and Domain-Freq. In this log-log plot, both analyses show a linear trend, indicating that the popularity of biomedical research resources follows a power-law distribution. This observation is consistent with a well-known phenomenon that very few individuals consume most resources and most individuals consume few resources.

\section{Service framework}

\begin{figure}
\centering
\includegraphics[width=0.7\linewidth]{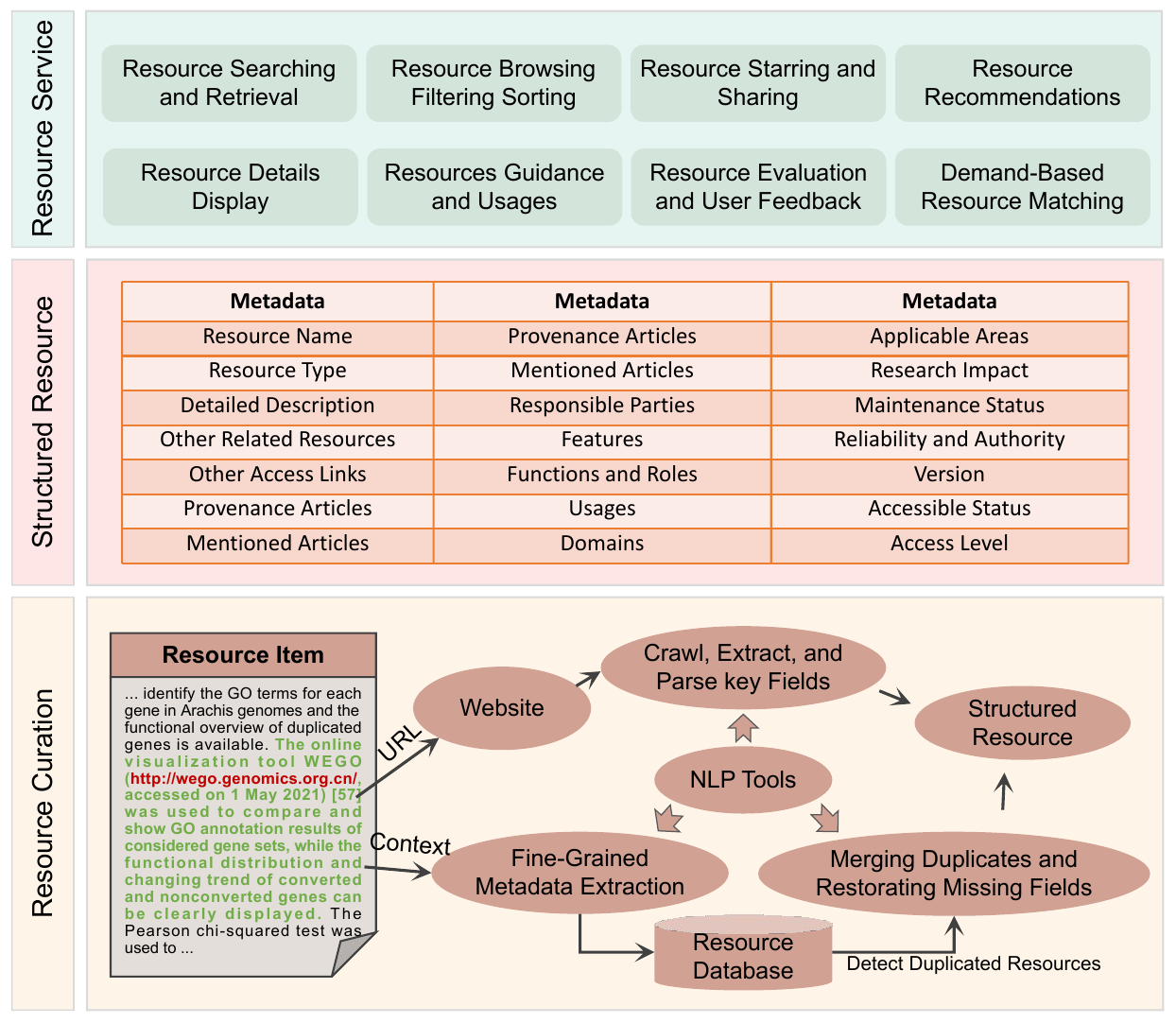}
\caption{Service framework}
\label{fig:service-framework}
\end{figure}

In this paper, we constructed a research resource database RRD-Bio for the biomedical field. The value of this database is that it centralizes the research resources in a single database, which provides a feasible solution for addressing the location and searching of research resources, as well as the information overload issue. However, the research is still in the early stage and cannot effectively serve researchers. There is a lot of work to be done, for example, processing the data to obtain machine-readable metadata such as resource name and type, and building a user-friendly system with some important functions integrated. To this end, this study designs a service framework, that aims to systematically improve the exploitation of the RRD-Bio database by building high-quality academic services based on the database. As shown in Fig.~\ref{fig:service-framework}, the framework includes how to curate resource data, what metadata needs to be extracted from resource items, and which functionalities should be built on top of the structured metadata.

Specifically, the data curation task aims to enhance resource description by extracting high-quality metadata from research resource entries using Natural Language Processing (NLP) techniques. Two data processing pathways have been designed. One involves locating the resource's webpage on the internet using the URL metadata and extracting comprehensive information from the webpage, including creator details, detailed resource descriptions, characteristics, user manuals, link availability, etc. The other pathway focuses on granular extraction based on the resource context, identifying resource type, function, source paper, referenced paper, functional roles, application areas, research impact, versions, and how authors utilize the resource. Additionally, since some resources may be mentioned multiple times in the literature, particularly popular tools, detecting the frequency of mentions and usage within the research community can contribute to enhancing the structured existence of resources in the database. Merging the extracted metadata from duplicated resources can enhance and complete missing metadata within the resource database.

Specifically, in data curation, we expect to extract some important metadata from research resources by using some NLP techniques, this will enhance the ability of resource description. Here, we design two data processing pipelines. The first one is to locate the resource's webpage on the Internet based on the URLs, and then obtain rich information from the webpages, such as the creator, features, user manual, and detailed description information of the source, and even the link is available. The second pipeline is to perform fine-grained extraction based on the extracted context of resources, such as identifying the resource type, creation papers, mentioned papers, functional roles, applicable research domains, academic impact of the resource, and how authors have used the resource. The complete list of metadata that can be extracted is shown in Fig.~\ref{fig:service-framework}. In addition, since some resources may be mentioned multiple times by many papers, such as popular research tools, the resources can be enhanced by other ``duplicate'' resources because the context of the resource may be significantly different. This can be achieved by detecting whether different resource mentions contain the same URLs.

The structured research resources facilitate the use of RRD-Bio. Based on this, we design a service framework for the RRD-Bio database, which contains the following important functionalities: resource searching and retrieval, resource details display, resource browsing filtering sorting, resources guidance and usages, resource starring and sharing, resource evaluation and user feedback, resource recommendations, demand-based resource matching. These functionalities are very important for maximizing the value of RRD-Bio, for example, for the users who plan to solve a new research problem, they can search the resource context to gain how others address similar problems, and what resources they used to address them. if the users don't find any related resources, they can use the browsing interfaces and resource recommendation function to obtain alternative resources. In addition, users can also view the details of the resource, including the creators, how the resource was used by others, the popularity of the resource, ranking the resource according to the latest mention time, starring the resources, sharing the resources, and rating the resources.

\section{Conclusions and limitations}

In this paper, we introduce a research resource database (RRD-Bio) for the biomedical field. This database comprises 2,555,116 resource mentions, collected from the two large biomedical scholarly databases, PMC and PubMed, which can be available from \url{https://zenodo.org/records/10526493}. We described the method of building this database, reported basic statistical indicators, and proposed a service framework to enhance the effective use of the database. In the future, we intend to improve the exploitation of the RRD-Bio database following the guidance of the service framework, such as extracting representative metadata, validating the accessibility of resource URLs, conducting survival analysis of resources, and building a service platform to support public use of the database.

There are several limitations in this study. One prominent limitation is that the resources in RRD-Bio are all derived from URL mentions in biomedical literature, ignoring those resources mentioned by literature citations, such as research papers published as datasets. This will reduce the coverage of biomedical research resources in RRD-Bio.

\bibliographystyle{unsrt}
\bibliography{main-pur-short}

\end{document}